\documentclass[a4paper,fleqn,usenatbib]{mnras}

\usepackage[T1]{fontenc}
\usepackage{ae,aecompl}
\usepackage{graphicx}	
\usepackage{amsmath}	
\usepackage{amssymb}	

\title{A novel look at the pulsar force-free magnetosphere}

\author[S. A. Petrova]{S. A. Petrova
\thanks{E-mail: petrova@rian.kharkov.ua}\\
Institute of Radio Astronomy, NAS of Ukraine, Chervonopraporna
Str., 4, Kharkov 61002, Ukraine}

\date{Accepted XXX. Received YYY; in original form ZZZ}

\pubyear{2016}

\begin{document}
\label{firstpage}
\pagerange{\pageref{firstpage}--\pageref{lastpage}}
\maketitle

\begin{abstract}
The stationary axisymmetric force-free magnetosphere of a pulsar is considered. We present an exact dipolar solution of the pulsar equation, construct the magnetospheric model on its basis and examine its observational support. The new model has toroidal rather than common cylindrical geometry, in line with that of the plasma outflow observed directly as the pulsar wind nebula at much larger spatial scale. In its new configuration, the axisymmetric magnetosphere consumes the neutron star rotational energy much more efficiently, implying re-estimation of the stellar magnetic field, $B_{\mathrm new}^0=3.3\times 10^{-4}B/P$, where $P$ is the pulsar period. Then the 7-order scatter of the magnetic field derived from the rotational characteristics of the pulsars observed appears consistent with the $\cot\chi$-law, where $\chi$ is a random quantity uniformly distributed in the interval $[0,\pi/2]$. Our result is suggestive of a unique actual magnetic field strength of the neutron stars along with a random angle between the magnetic and rotational axes and gives insight into the neutron star unification on the geometrical basis.
\end{abstract}


\section{Introduction}
The pulsar magnetosphere contains the copious electron-positron plasma which should affect the magnetospheric structure. In the axisymmetric ideal force-free model (where the pulsar rotation and magnetic axes are aligned and there is enough plasma to screen the accelerating electric field and to provide the electromagnetic force balance), the self-consistent distributions of the fields and currents are related by the non-linear pulsar equation \citep{m73,sw73,o74}. The only known exact solution of this equation \citep{m73} corresponds to the magnetic field of a monopole and is commonly believed to represent the magnetospheric structure far from the neutron star \citep[e.g.,][]{m91}. Consequently, in the numerically simulated magnetosphere \citep[e.g.,][]{ckf99,c05,s06,tsl13,b14,abinit14,gral14} the magnetic field varies smoothly from the dipolar form near the neutron star to the monopolar one at infinity, with the poloidal current distribution being close to that of a force-free monopole.

However, other exact solutions of the non-linear pulsar equation may have essentially distinct character and facilitate modification of the existing magnetospheric paradigm. Here we present an exact dipolar solution of the pulsar equation overlooked in the preceding literature. Both analytical and numerical studies of the pulsar equation were traditionally concentrated on the case of the magnetosphere corotating with the neutron star (or, as an exception, on the magnetosphere rotating at a different {\it constant} velocity \citep[e.g.,][]{c05}), mainly because there was no idea as to the actual physically grounded form of the rotational velocity distribution. It is well known that a pure dipole cannot be the solution of such a 'truncated' pulsar equation. In the present paper, we demonstrate that the 'complete' pulsar equation \citep{o74} does allow an exact dipolar solution with the unique poloidal current and rotational velocity distributions, construct the magnetospheric model on its basis and examine its observational support.

Due to the peculiar rotational velocity distribution, the purely dipolar force-free magnetosphere has toroidal rather than common cylindrical geometry, resembling the form of the pulsar wind nebula observed directly at much larger spatial scale \citep[see, e.g.,][for a review]{lyub15}. Besides that, the dipolar force-free magnetosphere appears substantially twisted and takes off the neutron star rotational energy much more efficiently not only as compared to the common magnetodipolar losses (of the orthogonal rotator) but also as compared to the losses of the classical force-free magnetosphere \citep{hard99}. This leads to a substantial re-estimation of the stellar magnetic field as function of the pulsar period. In particular, if the magnetars can be regarded as the axisymmetric force-free rotators, their magnetic field strength should be $\sim 10^{10}-10^{11}$ G, just as in the Central Compact Objects (CCOs). Furthermore, with our new magnetic field estimate, an unphysical trend of the stellar magnetic field strength with pulsar period is removed, whereas the residual scatter of the magnetic field values can be attributed to the random angle between the rotational and magnetic axes of a pulsar.

The plan of the paper is as follows. In Sect.~2, we present an exact dipolar solution of the pulsar equation, and its uniqueness is demonstrated in Appendix A. The dipolar force-free model of the pulsar magnetosphere is developed in Sect.~3. Observational consequences of our model are examined in Sect.~4. Section 5 contains a brief summary of the results.

\section{Pulsar equation and its exact solutions}
The presence of an abundant plasma in the pulsar magnetosphere implies that the magnetic and electric field strengths, $\bmath B$ and $\bmath E$, differ from those of a vacuum rotating dipole \citep[e.g.,][]{m73}. In case of an ideal conductivity $(\bmath{E}\cdot\bmath{B}=0)$, axial symmetry and stationarity, the fields can be presented as
\begin{equation}
\bmath{B}=\frac{\nabla f\times\bmath e_\varphi}{r\sin\theta}+\frac{A(f)}{r\sin\theta}\bmath{e}_\varphi, \quad \bmath{E}=-\Omega(f)\nabla f\,,
\end{equation}
where $(r,\theta,\varphi)$ are the coordinates in the spherical system with the axis along the pulsar axis, the dimensionless functions $f(r,\theta)$ and  $A(f)$ are proportional, respectively, to the magnetic flux and poloidal electric current through a circle of radius $\rho\equiv r\sin\theta$ centered on the magnetic axis at an altitude $z\equiv r\cos\theta$ above the origin, $\Omega(f)$ is the dimensionless angular frequency characterizing the differential rotation of the magnetosphere due to the drop of the electric potential $V(f)$ across the magnetic field lines, $\Omega\equiv\mathrm{d}V/\mathrm{d}f$.

With the dimensionless electric current and charge densities $\bmath{j}$ and $\rho_e$ given by the Maxwell's equations $\bmath{j}=\nabla\times\bmath{B}$ and $\rho_e=\nabla\cdot\bmath{E}$, the electromagnetic force balance, $\bmath{j}\times\bmath{B}+\rho_e\bmath{E}=0$, reduces to the pulsar equation \citep{o74}
\begin{equation}
\left (1-\Omega^2\rho^2\right )\Delta f-\frac{2}{\rho}\frac{\partial f}{\partial\rho}=-A\frac{\mathrm{d}A}{\mathrm{d}f}+\rho^2\Omega\frac{\mathrm{d}\Omega}{\mathrm{d}f}\left (\nabla f\right)^2\,,
\label{eq1}
\end{equation}	 	
where the functions $f$, $A(f)$ and $\Omega(f)$ are unknown. The classic monopolar solution has the form \citep{m73}
\begin{equation}
f=1-\cos\theta\,,\quad A=f(2-f)\,,\quad \Omega=1\,.
\label{eq2}
\end{equation}
It is easily generalized for an arbitrary $\Omega(f)$, in which case $A=\Omega f(2-f)$ \citep[see, e.g.,][]{p13}.

As is verified in Appendix, the pulsar equation (\ref{eq1}) also has a unique exact dipolar solution
	\begin{equation}
	f=\frac{\sin^2\theta}{r}\,,\quad \Omega=\frac{C_1}{f^2}\,,\quad A=\sqrt{\frac{4C_1^2}{f^2}+C_2},
	\label{eq3}
	\end{equation}	 	
where $C_1$ and $C_2$ are arbitrary constants. Among all conceivable solutions of Eq. (\ref{eq1}), it is the one seeming most relevant to the pulsar case.

At the magnetic axis $(f=0)$, the exact dipolar solution (\ref{eq3}) exhibits a singularity in both $A$ and $\Omega$, implying that in the axial region the force-free regime is broken. This contrasts with the monopolar case (\ref{eq2}), the consequent numerically simulated models \citep[e.g.,][]{ckf99,c05,s06,tsl13,b14,abinit14,gral14} and also with the numerical extensions toward the jet-like structure \citep{g04,l06,t14}, where  $A(0)$ and $\Omega(0)$ are finite, though the particle-in-cell simulations of the pulsar magnetosphere do arrive at a strong accelerating electric field along the magnetic axis \citep{ys12}.

\section{Basic features of the dipolar force-free magnetosphere}
\subsection{Alfvenic surface}
\begin{figure*}
\vspace{5cm}
\hspace{-10cm}
\includegraphics[width=125mm]{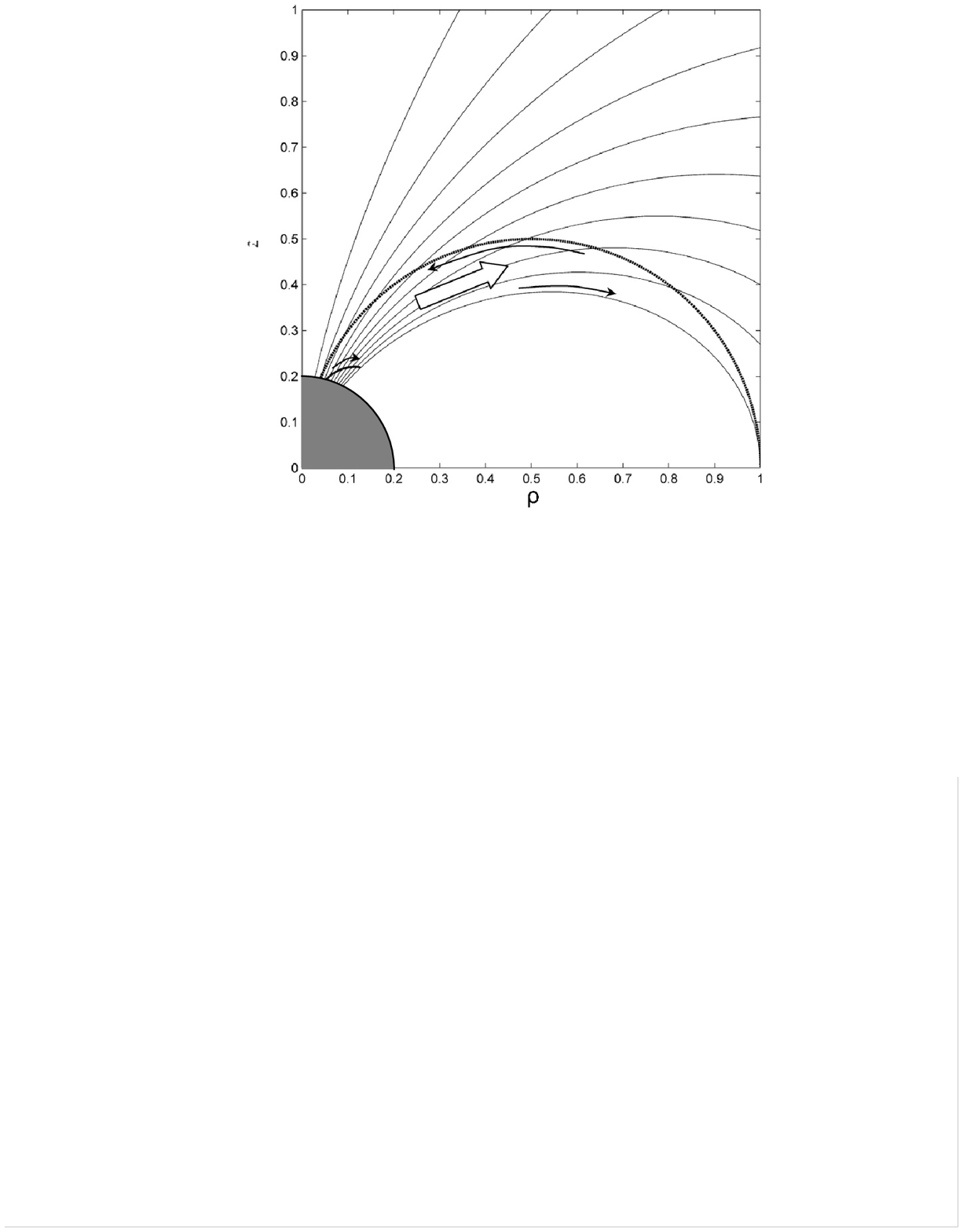}
\caption{The ideal force-free magnetosphere of an aligned rotating dipole: A general scheme; the magnetic field lines are shown by solid lines and correspond to the levels of $f$ from 0 to 1 with a step 0.1; the pair formation front is shown by the bold solid line, the light torus by the dashed line; the arrows mark the current flow directions.}
\label{f1}
\end{figure*}

The force-free magnetosphere described by Eq. (\ref{eq1}) has the peculiar surface at $\Omega r\sin\theta=1$, where the differential rotation velocity equals the speed of light. For constant $\Omega(f)$ it is known as the light cylinder, whereas in the purely dipolar case (\ref{eq3}) it is the torus, $r=C_1^{-1/3}\sin\theta$, whose large and small radii both equal to $C_1^{-1/3}/2$ (see Fig.~\ref{f1}).

The light torus separates the magnetosphere into the inner and outer parts, with the force-free regime holding only in the inner one. Then the poloidal current flowing along the force-free magnetic lines should close at the light torus with the consequent energy transmission from the electromagnetic field to the particles (see Sect.~3.2 below). The energy transmission at distances of order of the light torus size is supported by observations \citep{beskin10,ahar12}, and, in contrast to the cylindrical geometry, the toroidal one is appropriate for the current circuit closure. Leaving a place for an axial jet (at the dipolar magnetic field lines not crossing the light torus), the toroidal configuration of the inner magnetosphere resembles that observed in the pulsar wind nebulae at a much larger spatial scale \citep[e.g.,][]{lyub15}.

\subsection{Magnetospheric structure beyond the light torus}
Beyond the light torus, the force-free regime is expected to break, in which case the force-free electric field should switch off. According to the condition $\nabla\times\bmath{E}=0$, this should be followed by the rise of the longitudinal electric field component suggestive of particle acceleration and, perhaps, pair formation just beyond the light torus.

\begin{figure*}
\includegraphics[width=150mm]{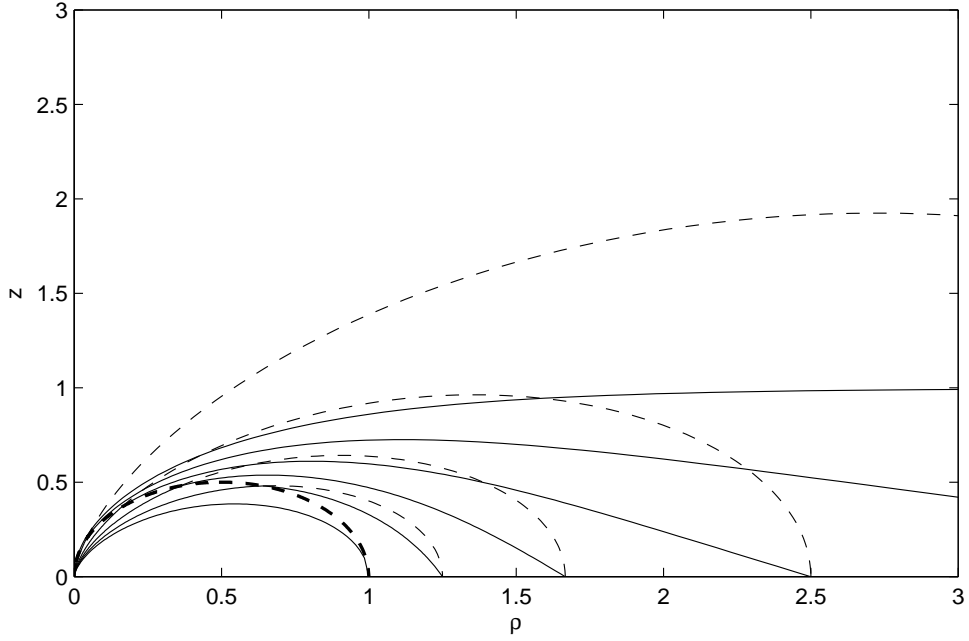}
\caption{The ideal force-free magnetosphere of an aligned rotating dipole: The region outside of the light torus; the light torus is shown by a bold dashed line, solid lines correspond to the magnetic field lines given by Eq.~\ref{eqa} with $a=1$, dashed lines display the field lines of a pure dipole, $f$ changes from 0.2 to 1 with a step 0.2.}
\label{f2}
\end{figure*}

Given that the azimuthal current beyond the light torus is absent, $(\nabla\times\bmath{B})_\varphi=0$, Eq.(\ref{a1}) is still valid and the poloidal magnetic field can be presented as a linear combination of the multipolar components. The magnetic flux continuity for the field lines crossing the light torus then yields
\begin{equation}
f=\frac{\sin^2\theta}{r}+\alpha\cos\theta-\alpha\cos\theta\frac{\sin^2\theta}{r^2}\,,
\label{eqa}
\end{equation}	 
where $\alpha$ is an arbitrary constant (see Fig.~\ref{f2} with $\alpha=1$). Independently of $\alpha$, the magnetic field lines meet the equator at $r=1/f$. The quantity $r_e=1/r_*$ (where $r_*$ is the dimensionless stellar radius) for the field line passing through the intersection of the light torus with the stellar surface presents one more magnetospheric scale.

\subsection{Appearance of the closed field line region}
\begin{figure*}
\includegraphics[width=120mm]{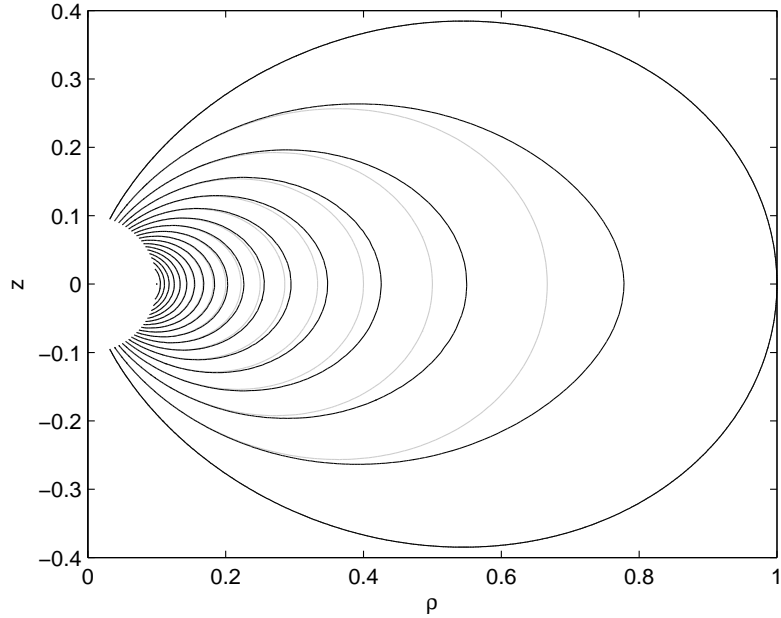}
\caption{The ideal force-free magnetosphere of an aligned rotating dipole: The closed field line region; solid lines represent the numerical force-free solution of the pulsar equation \ref{eq1} for $A\equiv 0$, $\Omega\equiv 1$ and the boundary condition $f=\sin^2\theta/r=1$, gray lines show the field lines of a pure dipole, the step in $f$ is 0.5.}
\label{f3}
\end{figure*}

Note that the dipolar solution (\ref{eq3}) allows arbitrary choice of the constants $C_1$ and $C_2$, in which case the poloidal current distribution may differ substantially. Hereafter we dwell on the case $C_1=1$, $C_2=-4$, which is most similar to what is commonly thought of a stable pulsar magnetosphere \citep[see, e.g.,][]{c05}. (The case $C_2=0$ corresponds to a corona-like structure resembling that suggested for the magnetars \citep{tlk02} and pulsars \citep{g06}). Then the closed field line region lies entirely inside the light torus and touches it at the equator (see Fig.~\ref{f3}). At the boundary of the closed region, $f=1$, we have $A=0$ and $\Omega=1$. A rigid corotation with the neutron star, $\Omega\equiv 1$, and the absence of a poloidal current, $A\equiv 0$, are also the typical attributes of the force-free closed region itself. The numerical solution of the pulsar equation (\ref{eq1}) in the closed force-free region with the boundary condition $f=\sin^2\theta/r=1$ is presented in Fig.~\ref{f3}.

\subsection{Physical picture of the current-carrying force-free region}
At $f<1$, the dipolar force-free magnetosphere appears substantially twisted. Although the poloidal magnetic field component dominates the azimuthal one even at the light torus, the twist is strong in a sense that the resultant electric charge and poloidal current densities $\rho_e$ and $j_p$ exceed the Goldreich-Julian ones, $\rho_\mathrm{GJ}$ and $\rho_\mathrm{GJ}c$ (where $\rho_\mathrm{GJ}\equiv B/Pc$ is the charge density necessary to screen the induction electric field of a vacuum rotating magnetic dipole \citep{gj69} and $c$ the speed of light):
\begin{equation}
\frac{j_p}{\rho_\mathrm{GJ}c}=\frac{2}{f^2\sqrt{1-f^2}}\,,\quad \frac{\rho_e}{\rho_\mathrm{GJ}}=\frac{2}{f^2}\frac{1+\sin^2\theta}{\sqrt{1+3\cos^2\theta}}\,.
\end{equation}	 

The super-Goldreich-Julian current density dictated by the dipolar force-free magnetosphere cannot be realized by the charges of one sign and therefore cannot be supplied directly from the neutron star surface. We suppose that at  $f<1$ the force-free region matches the vacuum one at the surface corresponding to the equality of the electric potentials. It is the surface where the closing current sheet should coexist with the pair formation front supplying the plasma necessary to sustain the force-free poloidal current \citep{p14}.

\subsubsection{Geometry of the pair formation front}
With the vacuum \citep{gj69} and force-free electric field potentials $\Phi$ and $V$ in the form
\begin{equation}
\Phi=-\frac{B_*\Omega_*R_*^5}{3cR^3}\frac{3\cos^2\theta-1}{2}\,,\quad V=-\frac{B_*\Omega_*^3R_*^3}{2c^3}\frac{R}{\sin^2\theta}
\end{equation}
(where $R\equiv rc/\Omega_*$ is the dimensional coordinate) the equation for the inner boundary of the force-free region reads
\begin{equation}
2\xi r^4=\sin^2\theta(3\cos^2\theta-1)
\end{equation}
with $\xi\equiv 3\Omega_*^2/2R_*^2c^2$. For $R_*=10^6$ cm we have $\xi\sim 10^8P^{-2}$ and $r\ll 10^{-2}P^{1/2}$; then the pair formation front lies close to the stellar surface, as in the common polar gap models. The two surfaces intersect at the polar angle $\theta_\mathrm{PFF}\approx\sqrt{\xi}r_*^2$, which is always larger than that of the light torus footprint at the stellar surface, $\theta_\mathrm{LT}\approx r_*$, with the difference, $\theta_\mathrm{PFF}-\theta_\mathrm{LT}\approx 0.225r_*$, presumably defining the actual width of the closing current layer as well as the size of the resultant hot spot at the stellar surface.

\subsubsection{Energy transfer in the current-carrying region}
In the regions encompassing the pair formation front (PFF) and the light torus (LT), the azimuthal magnetic field component can be presented as
\begin{equation}
B_{\varphi 1}=\frac{A\sigma(\mu_1)}{r\sin\theta}\,,\quad B_{\varphi 2}=\frac{A\left [1-\sigma(\mu_2)\right ]}{r\sin\theta}\,,
\end{equation}	 
where $\sigma(\mu)$ stands for the smoothed Heaviside function, whereas $\mu_1(r,\theta)$ and $\mu_2(r,\theta)$ are identically zero at the PFF and LT, respectively. Then the corresponding currents read
\[
\bmath{j}_1=\frac{\mathrm{d}A}{\mathrm{d}f}\frac{\nabla f\times\bmath{e}_\varphi}{r\sin\theta}\sigma(\mu_1)+\frac{A\delta(\mu_1)}{r\sin\theta}\left (\frac{\nabla\mu_1}{\vert\nabla\mu_1\vert}\times\bmath{e}_\varphi\right )\,,
\]	 	
\begin{equation}
\bmath{j}_2=\frac{\mathrm{d}A}{\mathrm{d}f}\frac{\nabla f\times\bmath{e}_\varphi}{r\sin\theta}\left [1-\sigma(\mu_2)\right ]-\frac{A\delta(\mu_2)}{r\sin\theta}\left (\frac{\nabla\mu_2}{\vert\nabla\mu_2\vert}\times\bmath{e}_\varphi\right )\,,
\end{equation}
where $\delta(\mu)\equiv\mathrm{d}\sigma/\mathrm{d}\mu$ is the smoothed Dirac delta-function. The first and second terms in the above equations correspond to the volume poloidal current and the closing current sheets, respectively.

The current sheets are not force-free, and the non-compensated azimuthal force is
\begin{equation}
F_{\varphi 1,2}=\pm\frac{A}{r^2\sin^2\theta}\left (\frac{\nabla\mu_{1,2}}{\vert\nabla\mu_{1,2}\vert}\times\nabla f\right )\,,
\label{eq4}
\end{equation}
where the plus sign corresponds to the subscript '1' and vice versa. The moment of the force (\ref{eq4}), $\bmath{K}\equiv \bmath{r}\times F_\varphi\bmath{e}_\varphi$, acts to spin up the PFF and to spin down the LT, adjusting the current sheets to their neighbourhood. The corresponding power, $-\bmath{K}\cdot\bmath{\Omega}$, taken off at the PFF and released at the LT, is
\begin{equation}
w=\frac{A\Omega}{r\sin\theta}\left (\frac{\nabla\mu_{1,2}}{\vert\nabla\mu_{1,2}\vert},\nabla f,\bmath{e}_\varphi\right )\,.
\label{eq5}
\end{equation}
The above equation appears exactly equal to the current losses in the sheets, $\bmath{j}_{1,2}\cdot\bmath{E}$, and also to the Poynting flux $(\bmath{E},\bmath{B},\nabla\mu/\vert\nabla\mu\vert)$ carried in the force-free region through the element of the revolving surface with the unit normal $\nabla\mu/\vert\nabla\mu\vert$.

Taking into account that $\bmath{e}_\varphi\times\nabla\mu/\vert\nabla\mu\vert$ is the unit vector tangent to the line $\mu=0$, $\bmath{e}_\tau\equiv\mathrm{d}\bmath{r}/\mathrm{d}\theta$, integration of Eq.(\ref{eq5}) over the whole surface is reduced to
\begin{equation}
W=\int{A(f)\Omega(f)\mathrm{d}f}
\end{equation}	 
and yields
\begin{equation}
W=\frac{\sqrt{1-r_*^2}}{r_*^2}-\log\frac{1+\sqrt{1-r_*^2}}{r}\,.
\end{equation}	 
As the dimensionless stellar radius is small, $r_*\ll 1$, the latter is simplified to $W\approx r_*^{-2}$. For the dimensional quantities, the surface integral of the power transmitted is
\begin{equation}
W\approx\frac{B_*^2\Omega_*^2R_*^4}{4c}\,,
\label{eq6}
\end{equation}
where $B_*$ is the magnetic field strength at the stellar surface, $\Omega_*$ the angular rotation frequency of the star, $R_*$ the stellar radius and both hemispheres are taken into account.

The assumed coexistence of the polar gap with the closing current sheet seems to have microphysical grounds \citep{p14}. The particles flowing across the poloidal magnetic field lines emit synchrotron photons, which are believed to be an important ingredient of the pair production scenario. The primary particles are accelerated in the longitudinal electric field resulting from the switching on of the force-free field on condition that $\nabla\times\bmath{E}=0$. Thus, it is the current losses that should ultimately provide the particle distribution sustaining the force-free regime at higher altitudes.

In the differentially rotating force-free region, the energy is transmitted along the magnetic field lines via the Poynting flux and is ultimately deposited to the particles at the light torus. The power transmitted is generally believed to be supplied by the current closing at the stellar surface and exerting a decelerating moment on the star \citep{bgi83}. In our scheme, however, most of the closing current sheet neighbours with the vacuum region, though a small segment does lie at the stellar surface. Apparently, it is the vacuum electromagnetic field that feeds the force-free region with energy and also adjusts the stellar rotation via interaction with its interior field.

\section{Applications to pulsars}
Leaving aside the details of energy transformation, we confront the power (\ref{eq6}) transmitted in the force-free region with the loss rate of the stellar rotational energy, $I\Omega_*\dot{\Omega}_*$ (where $I$ is the neutron star's moment of inertia and $\dot{\Omega}_*$ the temporal derivative of the stellar rotation frequency $\Omega_*$). The magnetodipolar losses of the neutron star,
\begin{equation}
W_\mathrm{dip}=\frac{2B_*^2\Omega_*^4R_*^6\sin^2\chi}{3c^3}
\label{eq7}
\end{equation}	 	
(where $\chi$ is the angle between the magnetic and rotation axes of a pulsar) are absent not only in the aligned rotator case $(\chi=0)$, since the radiation of a frequency $\sim\Omega_*$ should be formed over a region of a size $\sim c/\Omega_*$ actually occupied with the force-free plasma rather than vacuum \citep{bgi83}.

Comparison of Eqs. (\ref{eq6}) and (\ref{eq7}) proves that the aligned force-free magnetosphere can consume the stellar rotational energy much more efficiently than the vacuum orthogonal rotator $(\chi=\pi/2)$, with the surface magnetic field strength being re-estimated as
\begin{equation}
B_\mathrm{new}^0\approx 3.3\times 10^{-4}B/P\,,
\label{eq8}
\end{equation}
where $B=3.2\times 10^{19}(P\dot{P})^{1/2}$ G is the customary value based on magnetodipolar losses of the orthogonal rotator and it is taken that $I=10^{45}$ $\mathrm{g}\,\mathrm{cm}^{2}$ and $R_*=10^6$ cm. Thus, $B_\mathrm{new}^0$ appears much less, especially for long-period pulsars ($P>1$ s), though it is still within the limits expected for the neutron stars.

Interestingly, if the magnetars with their customary magnetic fields $B\sim 10^{14}-10^{15}$ G and periods $P=5-12$ s can be regarded as force-free aligned rotators, their actual fields should be $\sim 10^{10}-10^{11}$ G, just as those known for the CCOs \citep{mer15,kk16}. Because of the dramatic difference in the magnetic field strength, CCOs used to be regarded as 'anti-magnetars' \citep{halp07} or magnetars with the magnetic field buried under the crust \citep{vigano12}. Furthermore, it is the CCOs that do not exhibit any signature of the magnetospheric activity, being the most probable candidate for classical magnetodipolar losses, in which case the existing estimates of their magnetic field should remain true.

Note that the magnetar magnetic fields in the range $10^{14}-10^{15}$ G are generally believed to be directly inferred from the X-ray observations of cyclotron features \citep[see, e.g.,][]{tiengo13}. It should be kept in mind, however, that such estimates correspond to an assumption of proton lines. In case of electron lines, the magnetic field estimates should be less by a factor of $5\times 10^{-4}$, i.e. $\sim 10^{11}$ G, well in line with our result.

Our radical re-estimation of the magnetar magnetic field rules out the magnetic field energy as a dominant energy source of these objects. The magnetic field decay \citep{td96} is usually introduced  because the magnetar rotational energy loss rate is too low to account for the persistent X-ray luminosities observed, whereas their giant flares require the energy budget far in excess of the total stellar rotational energy. Note, however, that the X-ray luminosities of CCOs, with their low magnetic fields, also exceed their rotational energy loss rate. Perhaps, both types of neutron stars may be powered by the stellar thermal energy, which appears high enough to account for the magnetar energetics \citep{heyl05}.

A plethora of magnetar observational manifestations, including transient emission and drastic changes in the rotational characteristics \citep[see, e.g.,][for a recent review]{mer15}, hint at an unstable adjustment of the differentially rotating force-free region with its neighbourhood. Indeed, the neutron star deceleration implies an increase of the light torus, and this process does not necessarily hold monotonically.

\begin{figure*}
\includegraphics[width=100mm]{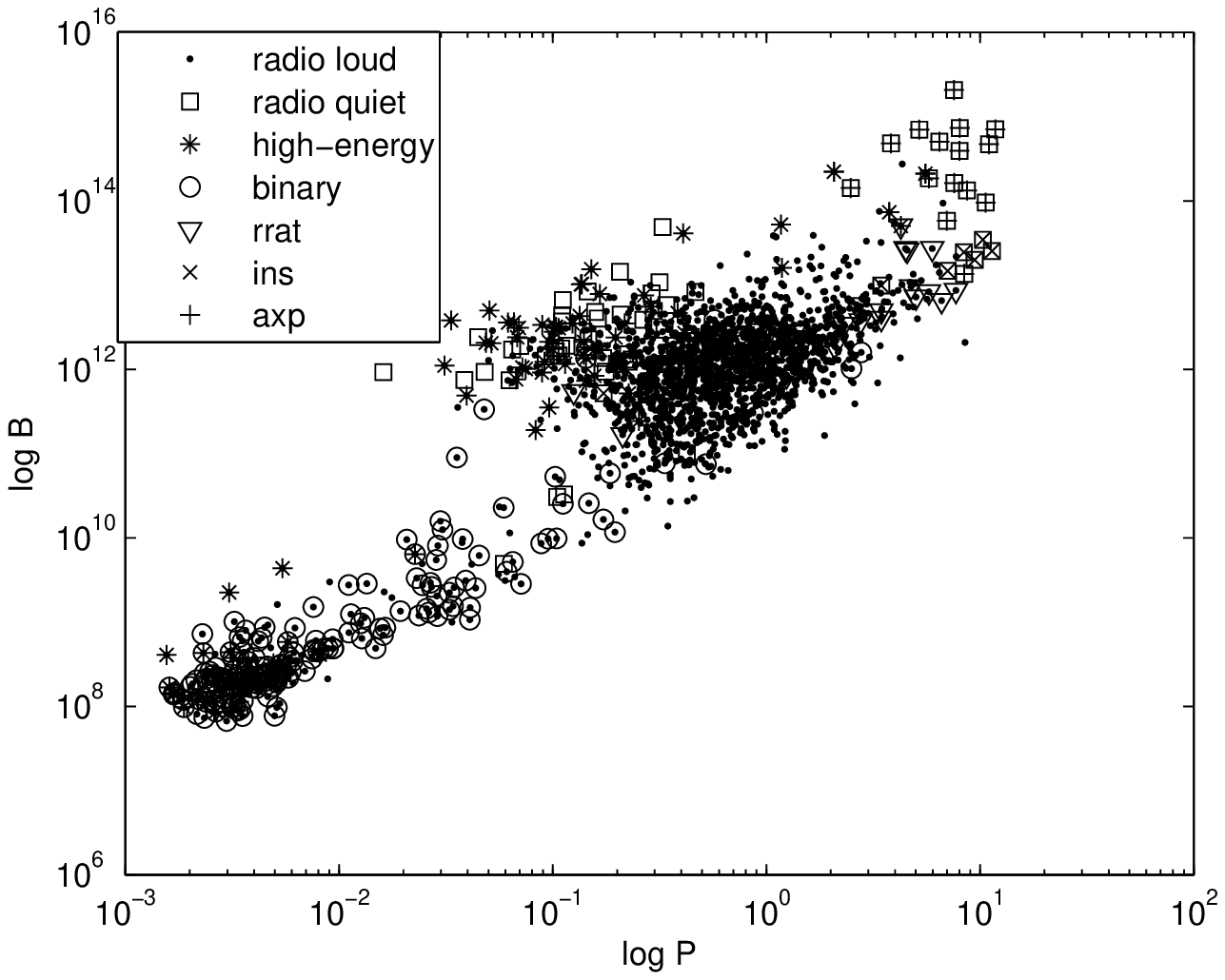}
\includegraphics[width=100mm]{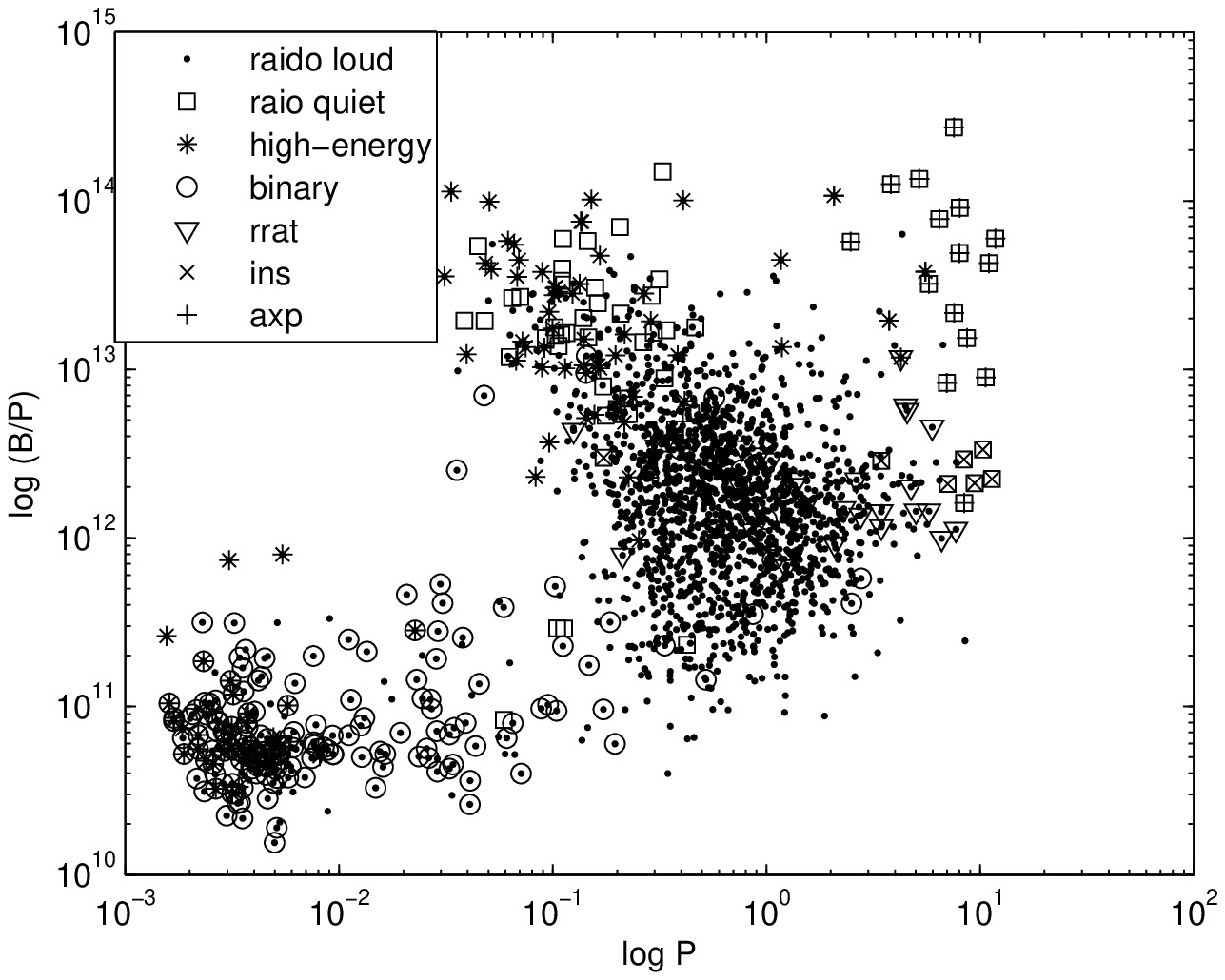}
\caption{Stellar magnetic field strength based on the pulsar rotational characteristics $P$ and $\dot{P}$ from the ATNF Pulsar Catalogue \citep{atnf05}; a) the case of magnetodipolar losses of an orthogonal vacuum dipole; b) the case of wind losses of an aligned force-free dipole.}
\label{f4}
\end{figure*}

Figures \ref{f4},a and \ref{f4},b show the magnetic field strength derived for the cases of the vacuum orthogonal and aligned force-free rotators as functions of the pulsar period for the pulsars from the ATNF Pulsar Catalogue \citep{atnf05}. The non-physical trend of $B$ with $P$ present in the former case is completely excluded in the latter one. In Fig. \ref{f4},b, the pulsars are expected to move with age toward longer periods at a constant $B/P$, approaching ultimately the death line. Take note of the sources with transient (RRATs) and quenched (INSs) radio emission near the expected death line.

The large and apparently random scatter of the points in Fig.\ref{f4},b is most probably attributable to the unaccounted non-zero angle $\chi$ in real pulsars. The distribution of the normal radio pulsars in $B/P$ appears accurately symmetric (in contrast to the distribution in $B$) and well fits the $\cot\chi$-law with $\chi$ being the random quantity uniformly distributed in the range $[0,\pi/2]$ (see Fig.\ref{f5} and Appendix B for details). Then the energy loss rate (\ref{eq6}) of an aligned force-free rotator can be empirically extended to the arbitrary inclination $\chi$ as
\begin{equation}
W\approx\frac{R_*^4\Omega_*^2(B_\mathrm{new}^0)^2}{4c}\frac{\cos^2\chi}{1+a^2\sin^2\chi}\,,
\end{equation}	 
where $a\gg 1$ is a constant and $B_\mathrm{new}^0$ is the same for all normal pulsars. At $\chi\to 0$ we arrive at the magnetar case, whereas at $\chi\to\pi/2$ the force-free losses cease, giving place to the magnetodipolar ones. With the maximum of the histogram in Fig.\ref{f5} at $\log B/P=12.25$, the normal pulsar magnetic field matches that of the CCOs and the re-estimated magnetar field on condition that $a\sim 10-10^2$.

\section{Conclusions}
We have found the exact dipolar solution of the pulsar equation and presented the magnetospheric model on its basis. With its general toroidal structure, the magnetosphere leaves a place for an axial jet, in striking contrast to the usual cylindric force-free models and in line with the jet+torus structure of the PWN observed.

The energy budget of the axisymmetric purely dipolar force-free magnetosphere appears much larger than that of the previous force-free models, implying substantial re-estimation of the neutron star magnetic field based on the pulsar rotational characteristics observed. In particular, if the magnetars can be regarded as the axisymmetric force-free rotators, their actual magnetic fields should be $\sim 10^{10}-10^{11}$ G, just like those of the CCOs, which are believed to lose rotational energy via magnetodipolar radiation and therefore to be characterized by the classical magnetic field estimates.

The new estimate of the neutron star magnetic field strength appears to have different dependence on the pulsar period. Consequently, an unphysical trend of the stellar magnetic field with period is removed. Furthermore, the residual scatter of the magnetic field values derived from the pulsar rotational characteristics may well be attributed to the random angle between the rotational and magnetic axes of pulsars. Then the magnetic field of neutron stars, similarly to their other basic parameters, such as the mass and radius, may be approximately the same for the whole neutron star population (perhaps, except for the millisecond pulsars with their peculiar evolution). As a result, unification of the neutron star properties \citep{kk16} may be realized primarily on the geometrical basis.

\section*{Acknowledgements}
The work have used the ATNF Pulsar Catalogue available at http://atnf.csiro.au/people/pulsar/psrcat


\begin{appendix}
\section{Exact dipolar solution of the pulsar equation}
For a pure dipole, $f=\sin^2\theta/r$, we have
\begin{equation}
\frac{\partial^2f}{\partial\rho^2}-\frac{1}{\rho}\frac{\partial f}{\partial\rho}+\frac{\partial^2f}{\partial z^2}=0
\label{a1}
\end{equation}
and therefore the pulsar equation (\ref{eq1}) can be presented in the form
\begin{equation}
F_1(f)\rho^2\Delta f+F_2(f)\rho^2(\nabla f)^2=F_3(f).
\label{a2}
\end{equation}
We have to find out whether the set of functions $F_{1,2,3}$ satisfying Eq.(\ref{a2}) exists and whether it is unique.

An arbitrary $F(\rho,z)$ is a function of $f$ given that the Poisson bracket,
\begin{equation}
\left [F,f\right ]\equiv\frac{\partial F}{\partial\rho}\frac{\partial f}{\partial z}-\frac{\partial F}{\partial z}\frac{\partial f}{\partial\rho}
\end{equation}	 
is zero. Applying the Poisson bracket to both sides of Eq.(\ref{a2}) yields
\begin{equation}
F_1(f)\left [\rho^2\Delta f,f\right ]+F_2(f)\left [\rho^2(\nabla f)^2,f\right ]=0\,,
\label{a3}
\end{equation}
which may be further reduced to
\begin{equation}
\left [\frac{\left [\rho^2\Delta f,f\right ]}{\left [\rho^2(\nabla f)^2,f\right ]},f\right ]=0.
\label{a4}
\end{equation}
For the purely dipolar flux function, Eq.(\ref{a4}) is fulfilled identically, and hence the set of functions $F_{1,2,3}(f)$ satisfying Eq.(\ref{a2}) really exists. Using Eqs.(\ref{a2}) and (\ref{a3}) and keeping in mind the actual form of the pulsar equation (\ref{eq1}), we arrive at the equations
\begin{equation}
\Omega^2=-\frac{f}{2}\Omega\frac{\mathrm{d}\Omega}{\mathrm{d}f}\,,\quad 2f^2\Omega\frac{\mathrm{d}\Omega}{\mathrm{d}f}=A\frac{\mathrm{d}A}{\mathrm{d}f}
\end{equation}
with a unique solution
\begin{equation}
\Omega=\frac{C_1}{f^2}\,,\quad A=\sqrt{\frac{4C_1^2}{f^2}+C_2}.
\label{a5}
\end{equation}

\section{Results of statistical analysis of the $B/P$ distribution for the normal radio pulsars}
\begin{figure*}
\includegraphics[width=120mm]{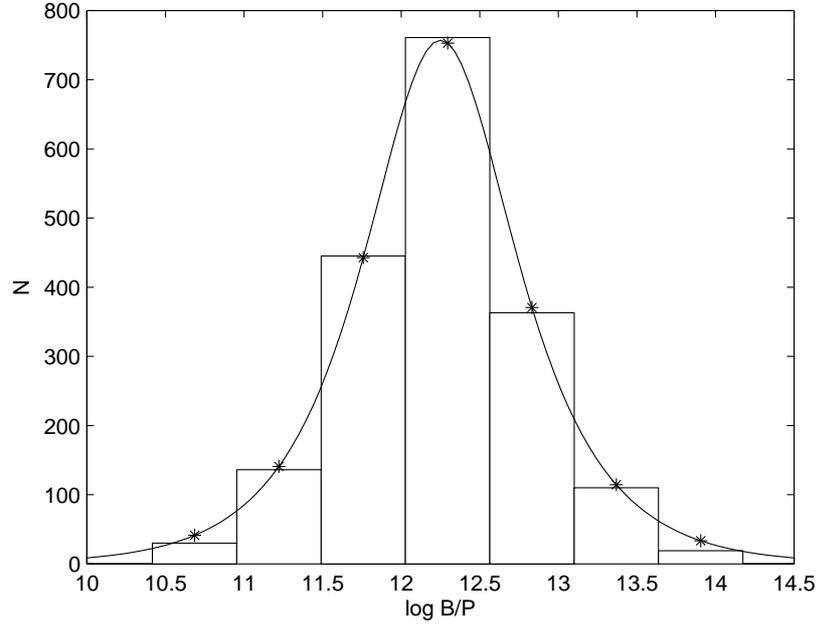}
\caption{Histogram of the force-free - based stellar magnetic field for the normal radio pulsars; the solid line with asterisks corresponds to the distribution of $\cot\chi$, where $\chi$ is the random quantity uniformly distributed in the interval $[0,\pi/2]$ (see Appendix B for the statistical analysis).}
\label{f5}
\end{figure*}

Based on Fig.~\ref{f5}, we hypothesize that the distribution of $B/P$ for the normal radio pulsars observed may be identified with the $\cot\chi$-distribution, where $\chi$ is the random quantity uniformly distributed in the interval $[0,\pi/2]$. According to the Pearson's criterium, this hypothesis cannot be rejected at a significance level 13\%. The Kolmogorov-Smirnov test proves that the two distributions are the same at a 88\%-significance level. For the 13-bin histogram of $B/P$ (not shown), the significance levels for the Pearson's criterium and Kolmogorov-Smirnov test are 5\% and 99.5\%, respectively.

\end{appendix}
\newpage

\newpage

\newpage

\newpage

\newpage



\bsp	
\label{lastpage}
\end{document}